\documentclass[twocolumn,showpacs,superscriptaddress,prb]{revtex4-1}
\usepackage{graphicx}
\usepackage{textcomp}
\usepackage{amsmath,amssymb,amsfonts,amsthm,latexsym} 
\usepackage[colorlinks=true,urlcolor=blue,linkcolor=blue,citecolor=blue,a4paper]{hyperref}
\usepackage{subfig}
\usepackage{gensymb}
\usepackage{setspace}
\usepackage{epstopdf}
\usepackage{xcolor,colortbl}
\usepackage{color}

\begin{document}
\title{Gate-tunable transport properties of in-situ capped Bi$_2$Te$_3$ topological insulator thin films}
\author{P. Ngabonziza}
\affiliation{Faculty of Science and Technology and MESA+ Institute for Nanotechnology, University of Twente, 7500 AE Enschede, The Netherlands}
\author{M. P. Stehno}
\affiliation{Faculty of Science and Technology and MESA+ Institute for Nanotechnology, University of Twente, 7500 AE Enschede, The Netherlands}
\author{H. Myoren}
\affiliation{Graduate school of Science and Engineering, Saitama University, 255 Shimo-Okubo, Sakura-ku, Saitama 338-8570, Japan}
\author{V. A. Neumann}
\affiliation{Faculty of Science and Technology and MESA+ Institute for Nanotechnology, University of Twente, 7500 AE Enschede, The Netherlands}
\date{\today}
\author{G. Koster}
\affiliation{Faculty of Science and Technology and MESA+ Institute for Nanotechnology, University of Twente, 7500 AE Enschede, The Netherlands}
\date{\today}
\author{A. Brinkman}
\affiliation{Faculty of Science and Technology and MESA+ Institute for Nanotechnology, University of Twente, 7500 AE Enschede, The Netherlands}
\date{\today}
\begin{abstract}
Combining the ability to prepare high-quality, intrinsic Bi$_2$Te$_3$ topological insulator thin films of low carrier density with in-situ protective capping, we demonstrate a pronounced, gate-tunable change in transport properties of Bi$_2$Te$_3$ thin films. Using a back-gate, the  carrier density is tuned by a factor of $\sim 7$ in Al$_2$O$_3$ capped Bi$_2$Te$_3$ sample and by a factor of  $\sim 2$ in Te capped Bi$_2$Te$_3$ films. We achieve full depletion of bulk carriers, which allows us to access the topological transport regime dominated by surface state conduction. When the Fermi level is placed in the bulk band gap, we observe the presence of two coherent conduction channels associated with the two decoupled surfaces. Our magnetotransport results show that the combination of capping layers and electrostatic tuning of the Fermi level provide a technological platform to investigate the topological properties of surface states in transport experiments and pave the way towards the implementation of a variety of topological quantum devices.
\end{abstract}
\pacs{73.50.-h, 73.20.-r, 75.47.-m}
\maketitle
\section{Introduction}
Three-dimensional (3D) topological insulators (TIs) such as HgTe, Bi$_2$Se$_3$, Bi$_2$Te$_3$, Sb$_2$Te$_3$, and their alloys,  are materials that are characterized by a narrow band gap in the electronic band structure while their surfaces host non-gapped spin-momentum locked states\cite{Kane2005,Fu2007, Zhang2009}. These helical surface states  are topologically protected owing to the strong spin-orbit coupling in the material, and they  mimic relativistic Dirac electrons \cite{Kane2005,Fu2007, Zhang2009}. TIs have generated much interest, not only as fundamentally new electronic states of matter, but also for their potential future technological applications in the fields of spintronics and quantum computation \cite{Hsieh2009,Moore2010,Jozwiak2013}.

Bismuth telluride (Bi$_2$Te$_3$) is a 3D TI material, which has been investigated extensively \cite{Chen2009,Alpichshev2010,Ngabonziza2015,Hoefer2014,Hoefer2015,Harrison2014,Virwani2014,Harrison2013}. The conducting surface states of Bi$_2$Te$_3$ have been observed first using surface-sensitive techniques like angle-resolved photoemission spectroscopy (ARPES) \cite{Chen2009} and scanning tunneling microscopy (STM), complemented by the scanning tunneling spectroscopy (STS) method \cite{Alpichshev2010}. Nevertheless, at present it remains challenging to find unambiguous evidence of surface state transport in experiments on Bi$_2$Te$_3$ single crystals or thin films. This is mainly because the charge transport is dominated by bulk conductivity due to residual carriers, which complicates the direct exploitation of the remarkable properties of the surface states.

Recently, progress has been made in synthesizing a new generation of high-quality, bulk-insulating Bi$_2$Te$_3$ thin films using molecular beam epitaxy\cite{Ngabonziza2015,Hoefer2014,Hoefer2015}. We have been able to grow Bi$_2$Te$_3$ thin films on different insulating substrates (STO [111] and Al$_2$O$_3$ [0001]) with the Fermi level placed in the bulk band gap\cite{Ngabonziza2015}. Combining different surface-sensitive in-situ characterization techniques, we demonstrated that the surface morphology and electronic band structure of Bi$_2$Te$_3$  are not affected by in-vacuo storage and exposure to oxygen, whereas major changes are observed when films are exposed to ambient conditions. Independently, combined in-situ four-point probe conductivity and angle resolved photoemission spectroscopy experiments revealed the bulk insulating properties of Bi$_2$Te$_3$ thin films with metallic surface states \cite{Hoefer2014,Hoefer2015}. In the work of Hoefer~\textit{et al.}\cite{Hoefer2014}, the authors also demonstrated that short exposure to air causes the bulk conduction band to be filled with electrons and the conductivity is no longer determined by the surface states alone. These observations highlight the need to cap Bi$_2$Te$_3$ thin films in-situ before exposing them to air and other ex-situ contaminations in order to protect the surface states from degradation and unintentional doping.

Different studies employed epitaxially-grown amorphous layers of Se/Te\cite{Harrison2014,Virwani2014} or an evaporated layer of Al \cite{Yu2013} as a protective capping layer for Bi$_2$Te$_3$ thin films whereas others used Al$_2$O$_3$ grown with atomic layer deposition (ALD) \cite{Liu2011}. However, it was noticed that the stoichiometry of the films was altered after removal of the Te/Se capping layer\cite{Harrison2014,Virwani2014} and that the ALD-grown Al$_2$O$_3$ layer can cause damage to the Bi$_2$Te$_3$ surface states\cite{Liu2011}. Such damage decreases the surface mobility and generates a large number of impurity states which pin the surface chemical potential, a situation that will complicate an efficient electrostatic tuning of the Fermi level \cite{Liu2011}. Only recently, elemental Te was identified as a suitable capping material for the protection of the topological surface of intrinsically insulating Bi$_2$Te$_3$ thin films\cite{Hoefer2015}, but systematic magnetotransport investigations of such capped films are needed. Further, an insulating capping layer is preferable because it allows for the fabrication of nearly damage-free top gate structures \cite{Yang2014}. Thus, it would be much more advantageous to develop a method for growing insulating capping layers with minimal damage to the surface states of intrinsically insulating Bi$_2$Te$_3$ thin films.

On the other hand, as the materials issues are being solved, the focus should now turn to finding an efficient way of controlling the Fermi level inside the bulk band gap. An effective gate control of topological surface states is highly desired for investigations of electronic transport properties in Bi$_2$Te$_3$ thin films and related device applications  \cite{Wiedenmann2016,Kurter2014,Yu2013,Liu2011,XHe2012,Yuan2011,Xiong2013}. 
To-date, uniform gate-modulation of the carrier density and chemical potential over large areas and wide ranges has not been demonstrated in (capped) Bi$_2$Te$_3$ samples. Our first demonstration of this capability in this work is an important step toward the implementation of large scale hybrid devices for quantum information or spintronics applications with Bi$_2$Te$_3$ thin films.

\begin{figure*}[!t]
{\includegraphics[width=1.\textwidth]{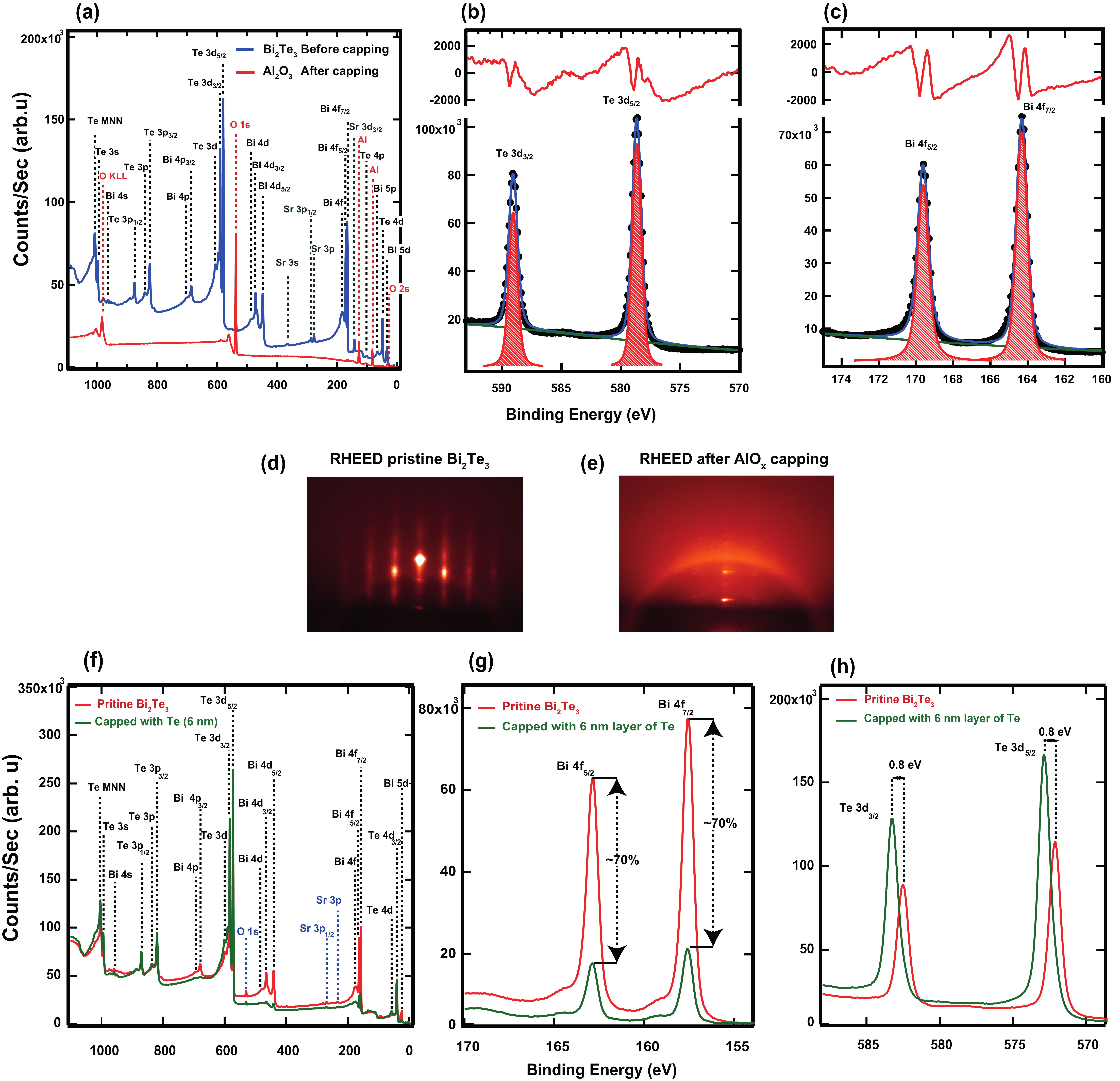}}  
  \caption{(Color online) In-situ capping and surface characterizations of 15 nm Bi$_2$Te$_3$ thin films. (a) An XPS survey scan taken after film growth (blue) and after capping  with a 6 nm Al$_2$O$_3$ layer (red). High-resolution scans around the Te $3d$ (b) and Bi $4f$ (c) main peaks, respectively, taken before capping in order to determine the surface stoichiometry. The red shaded zone shows the areas of the fitted peaks with the background (green line) removed. The upper red curve is the residual after Shirley background subtraction. RHEED patterns of the sample before (a) and after (b) in-situ capping with an insulating layer of Al$_2$O$_3$ grown by pulsed laser deposition. (f) XPS spectra of pristine Bi$_2$Te$_3$ (red) and after capping with an epitaxially grown 6 nm layer of Te (green). High resolution scans of Bi $4f$ (g) and Te $3d$ (h) core levels measured before and after capping.}
  \label{fig:SurfaceCheck}
\end{figure*}

We report bottom-gating results on in-situ capped, intrinsically insulating Bi$_2$Te$_3$ topological insulator thin films. We follow the growth recipe presented in Ref.~\cite{Ngabonziza2015} in order to obtain TI thin films with low densities of intrinsic carriers. First, we focus on determining the proper in-situ capping procedure of our Bi$_2$Te$_3$ thin films grown on different insulating substrates, namely Al$_2$O$_3$ and SrTiO$_3$. We use two different capping materials: an Al$_2$O$_3$ layer, grown by pulsed laser deposition (PLD), and an epitaxially grown Te layer. Secondly, we report a systematic magnetotransport characterization of the in-situ capped films by studying the bottom-gate dependence of the sheet resistance, 2D carrier density ($n_{\text{2D}}$) and weak antilocalization behavior. Here, we used the SrTiO$_3$ substrate as the back-gate dielectric. Lastly, we compare the transport characteristics of capped films with uncapped films of the same thickness and grown in the same conditions. Dual evidence from both, extracted carrier densities and weak antilocalization (WAL) phenomena, suggests that the Fermi level is in the bulk band gap for the in-situ capped films since capping prevents the formation of native oxide and isolates the film from different extrinsic defects \cite{Ngabonziza2015,Hoefer2014,Hoefer2015,Lang_M2012}. The transport data on in-situ capped Bi$_2$Te$_3$ thin films suggest that we can use the back-gate to tune the Fermi level 
all the way from the bulk conduction band (BCB) into the topological surface transport regime, and further down into the bulk valence band (BVB). These data reveal that capping the sample before taking it ex-situ for further electronic transport studies helps in preserving the surface states and avoiding unintentional doping induced by ambient conditions. 
\section{In-situ surface characterization}\label{FourTwo}
The high-quality Bi$_2$Te$_3$  thin films used in this work were grown by molecular-beam epitaxy (MBE) on insulating substrates, STO[111] and Al$_2$O$_3$[0001]. The growth procedure was similar to Ref.\cite{Ngabonziza2015}. Here, we compare magnetotransport properties of two samples of 15 nm thickness grown on STO. One sample was capped with amorphous Al$_2$O$_3$ and the other with an epitaxially-grown Te layer. The thickness of the protective capping layers of either material was $\sim$ 6 nm. For the purpose of understanding the role of capping, control samples without the protective layer were grown under the same conditions for comparison. Before and after capping, the surface condition of films were checked using X-ray photoemission spectroscopy (XPS) and reflection high energy electron diffraction (RHEED). For in-situ surface elemental characterization and detection of any Te/Bi excess in our films, we used XPS. Figure~\ref{fig:SurfaceCheck}\textcolor{blue}{(a)} shows the XPS spectra of the pristine (blue) film. Only the Bi and Te peaks are resolved with no appearance of extra peaks, confirming that the film is clean without contaminations. The analysis of the high resolution scans around the Te $3d$ and Bi $4f$ peaks for the pristine Bi$_2$Te$_3$ film (see Fig.~\ref{fig:SurfaceCheck}\textcolor{blue}{(b)} and \textcolor{blue}{(c)}) reveal narrow and symmetric core level lines, signaling the absence of Te or Bi excess in our film. The surface chemical stoichiometry was determined following the procedure presented in Ref.~\cite{Ngabonziza2015} and the Te:Bi ratio was determined to be 1.497$\pm 0.05$.  Figure~\ref{fig:SurfaceCheck}\textcolor{blue}{(d)} shows a typical RHEED pattern of the pristine Bi$_2$Te$_3$ film before being capped with an Al$_2$O$_3$ layer. The sharp RHEED streaks are indicative of a high-quality film with smooth single crystalline domains. After depositing the Al$_2$O$_3$ layer by pulsed laser deposition (PLD), the RHEED pattern of the same sample is shown in Fig.~\ref{fig:SurfaceCheck}\textcolor{blue}{(e)} for comparison. For the deposition, the films had been transferred to the PLD chamber without breaking high vacuum (HV) conditions as the PLD system is connected to the growth chamber via an HV distribution chamber. The capping layer was deposited at room temperature in argon flow (40 sccm) at a background pressure of 1$\times10^{-1}$ mbar using a laser energy density of 2.75 J/cm$^2$. In order to minimize the degradation of the Bi$_2$Te$_3$ surface states due to high energy ionic bombardment of the sample surface during the PLD process, the sample holder was placed at a relatively far distance ($\sim 60$ mm) from the target material. After the Al$_2$O$_3$ deposition, only the oxygen and aluminum peaks were resolved in the XPS spectrum (red curve in Fig.~\ref{fig:SurfaceCheck}\textcolor{blue}{(a)}).

For samples capped with Te, the capping was also performed at room temperature, immediately after the growth of Bi$_2$Te$_3$ thin films in the MBE growth chamber without breaking the ultra-high vacuum conditions. The surface elemental composition of these films was also investigated using XPS. Only Bi and Te peaks are resolved in the XPS spectra before and after the deposition of the 6 nm Te capping layer (see Fig.~\ref{fig:SurfaceCheck}\textcolor{blue}{(f)}). The intensities of the Bi core levels in Bi$_2$Te$_3$ are observed to decrease after the capping step. From the analysis of the high resolution XPS spectra (see Fig.~\ref{fig:SurfaceCheck}\textcolor{blue}{(g)} and \textcolor{blue}{(h)}), the Bi $4f$ spectral intensity is determined to have reduced by $\sim$ 70\%  without any changes in the lineshape or energy position. This indicates that the Bi in Bi$_2$Te$_3$ is not chemically affected by the deposited 6 nm capping layer of Te\cite{Hoefer2015}. On the other hand, the Te $3d_{5/2}$ and Te $3d_{3/2}$ peaks are slightly influenced by this capping procedure since their lineshape peak positions are shifted by $\sim$ 0.8 eV toward higher binding energies upon the deposition of the Te overlayer. This shoulder has been observed previously and was attributed to elemental Te \cite{Hoefer2015,Moulder1996}. The XPS data on Te capped film are in agreement with recent in-situ ARPES and four-point probe conductivity data on Bi$_2$Te$_3$ films \cite{Hoefer2015}, which revealed that the surface states of the pristine Bi$_2$Te$_3$ samples are not affected by the Te capping, thus suggesting the Te capping method does not cause a doping of the sample.

\section{Gate-tunable transport characteristics}\label{FourThree}

\begin{figure*}[!t]
{\includegraphics[width=0.925\textwidth]{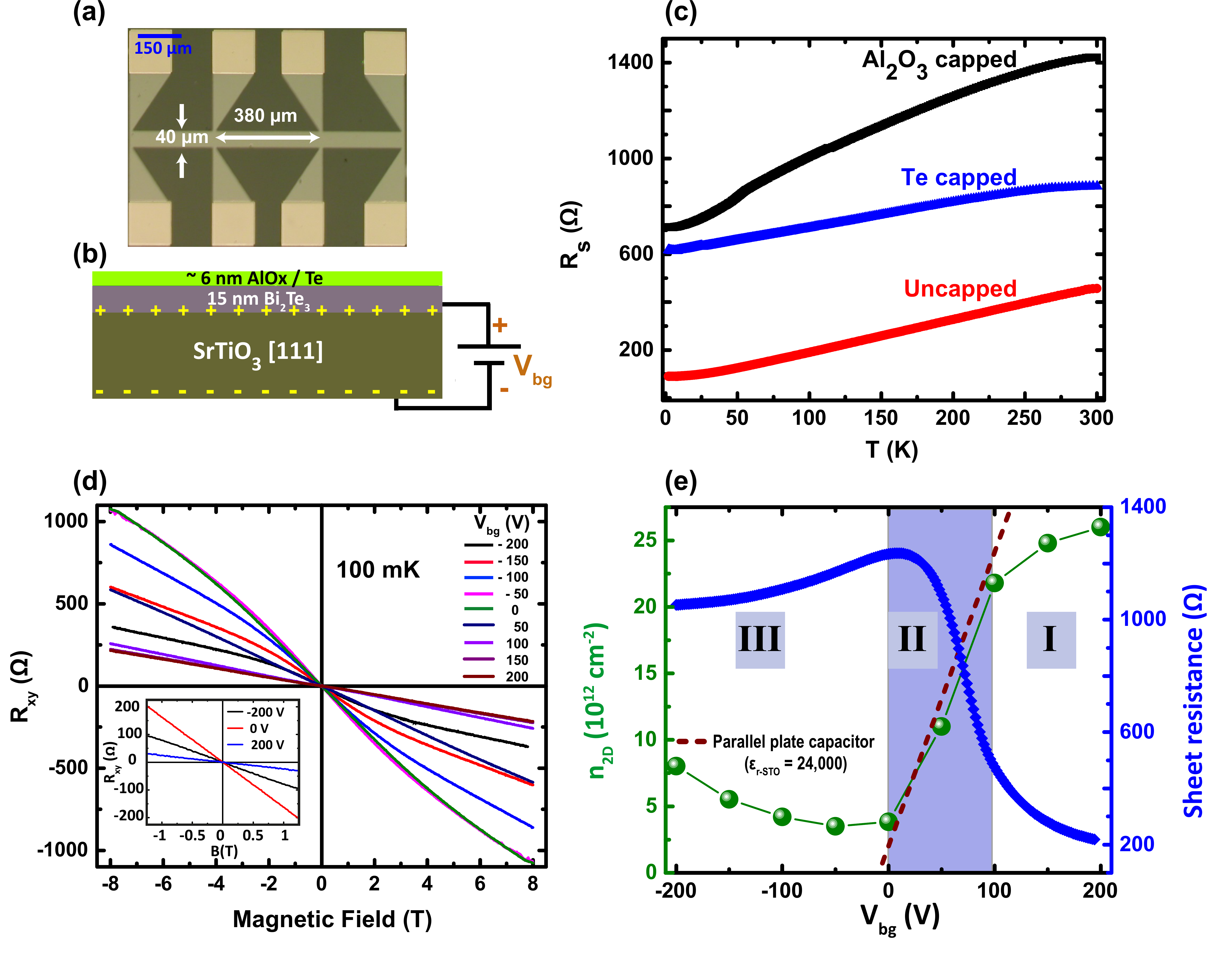}}  
  \caption{(Color online) Transport characteristics of 15 nm Bi$_2$Te$_3$ thin films. (a) Optical micrograph of the structured Hall bar device. (b) Schematic representation of a typical in-situ capped Bi$_2$Te$_3$ film grown on SrTiO$_3$ [111]. The substrate is used as the back-gate dielectric in transport experiments. The total thickness of the capping layers (of either Al$_2$O$_3$ or Te) is $\sim 6$~nm. (c) Temperature dependence of the zero-field sheet resistance for three different Bi$_2$Te$_3$ films. Black: capped with Al$_2$O$_3$, blue: capped with Te and red: uncapped film. (d) Back-gate voltage dependence of the Hall resistance, $R_{\text{xy}}$ at 100 mK. (e) Back-gate voltage ($V_{bg}$) dependence of the 2D carrier density extracted from the Hall zero-field slope of $R_{\text{xy}}$ in (d) and of the sheet resistance. Three different gate voltage regions I, II and III are identified with distinct behavior. The change in 2D carrier density follows the parallel plate capacitor law in the intermediate $V_{bg}$ regime (dashed line fit in region II).}
  \label{fig:TransportCharacteristics}
\end{figure*}

In order to investigate the electronic transport properties of capped films and to study the effect of the protective capping layer in magnetransport properties of in-situ capped and uncapped Bi$_2$Te$_3$ thin films, we fabricated Hall bar devices using standard photolithography methods and subsequent wet etching. For the etching step, we used a solution of nitric acid (HNO$_3$) diluted with de-ionized water (1 HNO$_3$ :3 H$_2$O). Gold contacts were added in a subsequent lithography step. To achieve proper electric contact in the Al$_2$O$_3$ capped samples, we removed the insulating layer in the contact area by etching the aluminum oxide with the same developer solution (a base) that we use for photolithography. Figure~\ref{fig:TransportCharacteristics}\textcolor{blue}{(a)} shows an optical image of a typical Hall bar device. We used the STO [111] substrate as the back-gate dielectric since it has a high dielectric constant at low temperature ($\sim 24 000$). The same approach has already been used for Bi$_2$Se$_3$ and (Bi$_{0.5}$Sb$_{0.5}$)$_2$Te$_3$ thin films, and Bi$_2$Te$_3$ nanoribbons also using STO [111] as the back-gate dielectric\cite{Zhang2011,XHe2012,JChen2011,JChen2010,AJauregui2015,FXiu2011}. Figure~\ref{fig:TransportCharacteristics}\textcolor{blue}{(b)} depicts a schematic illustration of the bottom-gated thin film. Electrical transport properties of the films were first characterized in the temperature range from $300\, $K down to $2\, $K using a Quantum Design PPMS measurement system. Later, we conducted magnetotransport measurements at different back-gate voltages in a dilution refrigerator at 100 mK. The measurements were performed using a standard lock-in technique with a small excitation current of 5 nA. Both, the sheet resistance $R_\text{S}$ and the Hall resistance $R_{\text{xy}}$, were measured as a function of magnetic field at different back-gate voltages ($V_{bg}$), in the range of $\pm 200\,$V. Sweeping the gate voltage back and forth, we observed hysteretic behavior,  probably due to the presence of impurity states that trap charges as has been suggested by previous studies\cite{Yang2014,XHe2012,JTian2014}. In order to keep consistency in the data, all graphs presented here were taken by sweeping $V_{bg}$ from positive to negative values.

\begin{figure*}[!t]
{\includegraphics[width=0.925\textwidth]{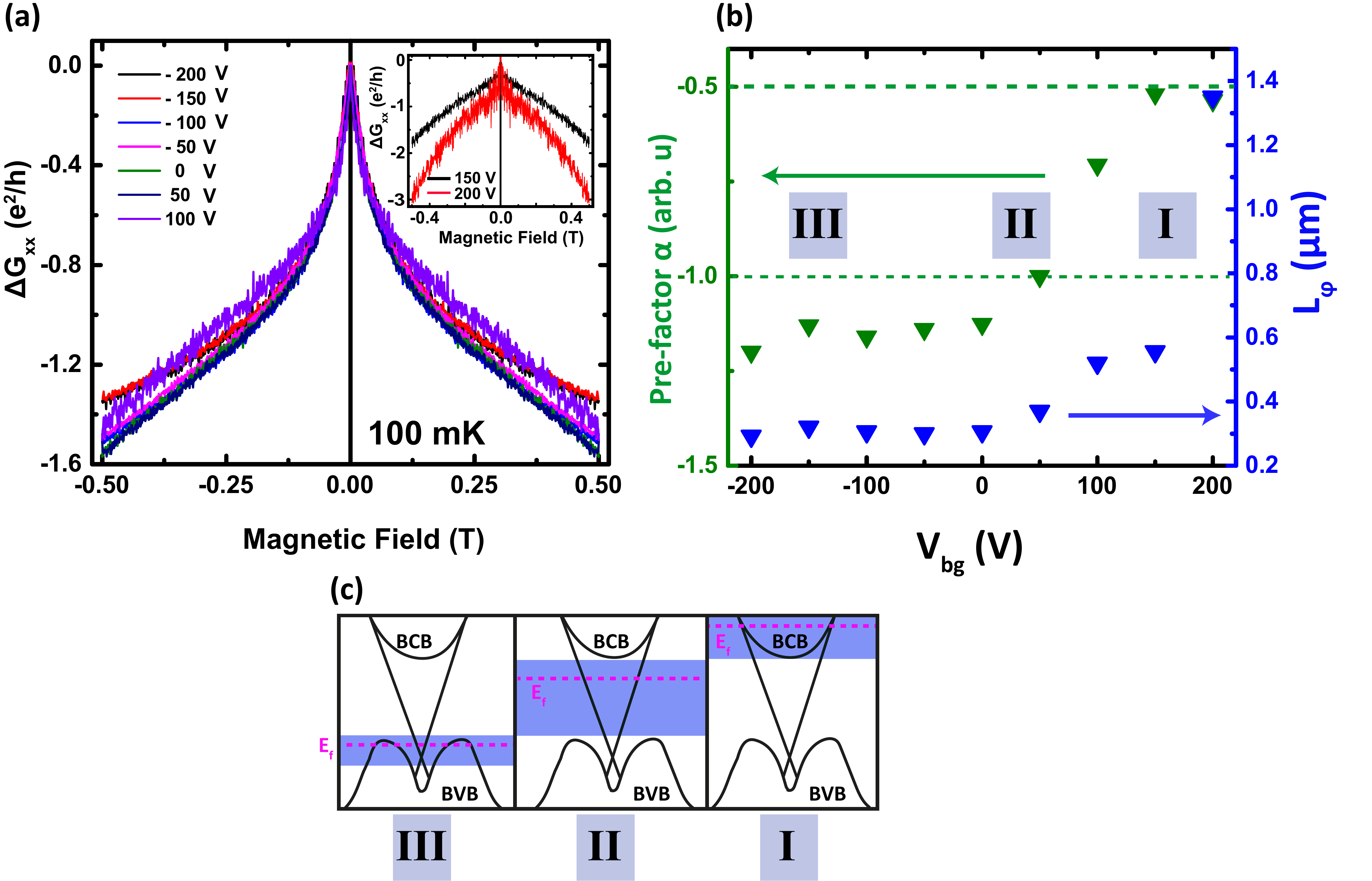}}  
  \caption{ Back-gate dependence of the WAL effect of an in-situ Al$_2$O$_3$-capped Bi$_2$Te$_3$ thin film. (a) The gate-voltage dependence of the low field magnetoconductance $\Delta G_\text{xx}(B)$. All data were taken at 100 mK. The inset highlights the observation of a parabolic background contribution to $\Delta G_\text{xx}$ at high positive $V_{bg}$. (b) Gate dependence of the prefactor $\alpha$ and the dephasing length $l_{\varphi}$, obtained from best fits to equation \textcolor{blue}{(\ref{HLN})}. (c) A schematic band diagram illustration relevant to the thin films used in this study. BCB is the bulk conduction band, BVB is the bulk valence band, and $E_{f}$ is the position of the Fermi level. We access the transport regimes I, II and III of (b) as the Fermi level moves through the respective shaded regions of the band structure in (c) and Fig.~\ref{fig:TransportCharacteristics}\textcolor{blue}{(e)} when the gate voltage is lowered.}
  \label{fig:WALbehaviour}
\end{figure*}

Figure \ref{fig:TransportCharacteristics}\textcolor{blue}{(c)} presents the temperature-dependent zero magnetic field sheet resistance for two different capped Bi$_2$Te$_3$ films (black: capped with Al$_2$O$_3$, blue: capped with Te), and for uncapped film (red curve). All three samples show metallic behaviour over the entire temperature range; but the sheet resistance is lower in the uncapped film by about $500\,\Omega$ and $ 600\,\Omega$ at 2 K compared to the Te and Al$_2$O$_3$-capped films, respectively. This observation is consistent with recent in-situ ARPES and four probe STM transport measurements on pristine, capped and exposed to air Bi$_2$Te$_3$ thin films \cite{Hoefer2014,Hoefer2015}. The decrease in $R_\text{S}$ was attributed to originate from exposing films to ambient conditions; and corresponding ARPES spectra on the same pristine and capped films proved that the Fermi level is inside the bulk band gap prior and after in-situ transport measurements, which always showed metallic behaviour of the Bi$_2$Te$_3$ thin films\cite{Hoefer2014,Hoefer2015}.

\subsection{Back-gate dependence of in-situ Al$_2$O$_3$-capped Bi$_2$Te$_3$ film}
We first discuss the magnetotransport properties of the Al$_2$O$_3$-capped Bi$_2$Te$_3$ thin film. The slope of $R_{\text{xy}}$ indicates $n$-type carriers.  It increases with decreasing $V_{bg}$. The  $R_{\text{xy}}$ curves start to show  nonlinearity below $V_{bg}=50$V (see Fig.~\ref{fig:TransportCharacteristics}\textcolor{blue}{(d)}). Similar behavior has been reported in previous work on Bi$_2$Se$_3$ and (Bi$_{1-\text{x}}$Sb$_{\text{x}}$)$_2$Te$_3$ thin films grown on SrTiO$_3$ [111] substrate \cite{Zhang2011,XHe2012,JChen2011,JChen2010}. 
We do not observe a reversal of the sign of $R_{\text{xy}}$ in the entire magnetic field range in contrast to, e.g., (Bi$_{1-\text{x}}$Sb$_{\text{x}}$)$_2$Te$_3$ samples in Ref.~\cite{XHe2012}. In the case of Bi$_2$Se$_3$ thin films grown on STO [111], the change of the sign in the Hall signal has only been achieved at much higher magnetic fields and for large negative back-gate voltages~\cite{JChen2011,Zhang2011}. We conjecture that the ambipolar effect is not observed within the accessible back-gate field range because of the ``M-like'' shape of the bulk valence band (BVB) structure of Bi$_2$Te$_3$~\cite{Chen2009,Ngabonziza2015} (see Fig.~\ref{fig:WALbehaviour}\textcolor{blue}{(c)} and related discussion below). The Dirac point (DP) is buried inside the BVB, thus only $n$-type topological surface states carriers are accessible inside the bulk band gap, as it has also been recently suggested from gate-tunable transport studies of Bi$_2$Te$_3$ nanoribbon \cite{AJauregui2015}.

The nonlinearity in the Hall resistivity curves below 50\,V suggests the coexistence of multiple charge carrier types. We extracted the low-field Hall coefficients ($R_H$) which are calculated from the slope of the $R_{\text{xy}}$ curves for $B<1.5\,$T (see inset Fig.~\ref{fig:TransportCharacteristics}\textcolor{blue}{(d)}).  It should be emphasized here that, in the case where $R_{\text{xy}}$(B) is nonlinear, a multiple band model approach should be taken. However, the discrepancies in carrier density and mobility of the high-mobility carriers (corresponding to the low-field Hall coefficient)  are small between one- and two-band fits to $R_{\text{xy}}$. We were not able to obtain a simultaneous fit to both components of the magnetoresistance data, $R_{\text{xx}}$ and $R_{\text{xy}}$, within the framework of the standard two-band model. A full theory of band structure and quantum corrections to the conduction of 3D TI surface and bulk bands is still missing. Thus, the 2D carrier density ($n_{2D}$) is extracted from $n_{2D}=\frac{1}{e\, R_H}$, with the low-field Hall coefficient $R_H$ as defined above, and $e$ the electronic charge. The mobility is calculated using $\mu=\frac{1}{e\, R_s n_{2D}}$ at zero magnetic field.

The response of sheet carrier density and sheet resistance of the Al$_2$O$_3$-capped Bi$_2$Te$_3$ thin film to the applied back-gate voltage is summarized in Fig.~\ref{fig:TransportCharacteristics}\textcolor{blue}{(e)}. The first notable observation is the appearance of a maximum in the sheet resistance at V$_{bg}\sim 10\, $V (region II), which is the first time, to our knowledge, that this behavior is seen in Bi$_2$Te$_3$. Previously, it has only been observed in Bi$_2$Se$_3$ and (Bi$_{1-\text{x}}$Sb$_{\text{x}}$)$_2$Te$_3$ single crystals and thin films \cite{JChen2010,JChen2011,Zhang2011,XHe2012,JTian2014,JLee2012}, and Bi$_2$Te$_3$ nanoribbons placed on a SrTiO$_3$ substrate \cite{AJauregui2015}. 
However, we would like to point out that the resistance maximum is not correlated with the depletion of the surface states at the Dirac point (DP), cp. Refs.~\cite{JChen2010,JChen2011,Zhang2011,JLee2012}, as the DP is buried below the BVB maximum. We suggest that the downturn and saturation of the sheet resistance at negative gate voltages (region III) is similar to what was observed in the measurements by Lee \textit{et al.} \cite{JLee2012} on Bi$_{1.5}$Sb$_{0.5}$Te$_{1.7}$Se$_{1.3}$ single crystals. The results were interpreted as an increase in interband scattering when the Fermi level touches the bulk valence band (see region III in Fig.~\ref{fig:WALbehaviour}\textcolor{blue}{(c)}).

In total, we modulate the carrier density $n_{\text{2D}}$ by a factor of $\sim 7$ in the accessible back-gate voltage range ($\pm 200~V$). The biggest drop in carrier density is seen in region II of Fig.~\ref{fig:TransportCharacteristics}\textcolor{blue}{(e)}. 
The lowest extracted carrier density is $n_{\text{2D}}= 3.5\times 10^{12}\text{cm}^{-2}$ at $V_{bg}\sim -50\,$V with a corresponding mobility of $1,600\ \text{cm}^2\, V^{-1} s^{-1}$.
The low carrier concentration in the intermediate region strongly suggests that the Fermi level is located in the bulk band gap, crossing only topological surface states. This is further corroborated by weak antilocalization measurements presented later. Furthermore, $n_{\text{2D}}$ versus $V_{bg}$ is observed to be linear as one would expect from a parallel plate capacitor model (dashed line fit in Fig.~\ref{fig:TransportCharacteristics}\textcolor{blue}{(e)}). All added carriers participate in electric conduction in the intermediate $V_{bg}$ regime, a behavior which is expected for surface states.

At large, positive and at negative back-gate voltages, the linear relationship does not hold. In region III, the carrier density starts to increase slowly again as the bulk valence band contributes mobile carriers. The Dirac point is buried deeper in the bulk valence band in Bi$_2$Te$_3$ samples which makes it unaccessible by electrostatic gating in the gate voltage range of our experimental setup\cite{Chen2009,Ngabonziza2015,Hoefer2014}. In region I, we add fewer high-mobility carriers by increasing the gate voltage. We attribute this to the onset of bulk conduction when the Fermi level crosses the bottom of the bulk conduction band (BCB).

Additional information about the carrier distribution and the scattering between transport channels is obtained from a measurement of the weak (anti-)localization effect\cite{Hikami1980}. The WAL effect in TIs has been  studied intensively both theoretically \cite{HZLu2011,IGarate2012} and experimentally \cite{Yang2014,JChen2011,JChen2010,Chiu2013}. WAL is a quantum effect that is commonly observed in TIs due to the spin-momentum locking resulting from strong spin-orbit coupling\cite{HZLu2014}. The $
\pi$-Berry's phase of the topological surface states leads to destructive interference between time-reversed charge carrier paths, thus increasing the conductivity at zero magnetic field. The cusp-like WAL maximum can be destroyed by applying an external magnetic field to the sample, which breaks the time reversal symmetry and $\pi$-Berry's phase\cite{HZLu2014,HZLu2011}. Figure \ref{fig:WALbehaviour}\textcolor{blue}{(a)} gives the sheet conductance $\Delta G_\text{xx}(B)= G_\text{xx}(B)-G_\text{xx}(B=0\, \text{T})$ as a function of magnetic field for various $V_{bg}$ measured at 100 mK. The $\Delta G_\text{xx}(B)$ data show sharp cusp-like maxima at zero magnetic field, indicating the existence of the WAL. At high positive $V_{bg}$, the WAL effect is suppressed, containing parabolic background (see inset Fig.~\ref{fig:WALbehaviour}\textcolor{blue}{(a)}). The suppression of WAL at high $V_{bg}$ can be due to either enhanced bulk conductivity when $E_f$ is located in the bulk or disorder induced electron-electron interactions in the system \cite{Chiu2013,JWang2013}. The former scenario is much more likely in our samples since it is consistent with the $n_{2D}$ analysis presented in Fig.~\ref{fig:TransportCharacteristics}\textcolor{blue}{(e)} and is similar to the behaviors reported previously in gating experiments \cite{JTian2014,Steinberg2011}. 

The WAL behaviour is described quantitatively using the simplified Hikami-Larkin-Nagaoka (HLN) equation\cite{Hikami1980, Maekawa1981}. In the limit of strong spin-orbit coupling ($\tau_{\varphi}\gg \tau_{so},\tau_e$; where $\tau_{\varphi}$ is the dephasing time, $\tau_{so}$ the spin-orbit scattering time, and $\tau_e$ the elastic scattering time), with a negligible Zeeman term, the magnetoconductance correction
is given by:
 \begin{equation}\label{HLN}
 \Delta G_\text{xx}(B)= -\alpha \frac{e^2}{\pi h} \left[ \varPsi \left( \frac{1}{2}+\frac{B_{\varphi}}{B}\right)-\ln \left(\frac{B_{\varphi}}{B}  \right) \right];
 \end{equation}
 where $\alpha$ is a parameter whose value reflects the number
of conduction channels. $\varPsi$ is the digamma function, $B_{\varphi}= \hbar/4e\,l^2_{\varphi}$ is the characteristic magnetic field with $l^2_{\varphi}=D\tau_{\varphi}$, which is the phase coherent length, and $D$ the diffusion constant. The relationship between the prefactor $\alpha$ and the number of conducting channels in equation~\textcolor{blue}{(\ref{HLN})} is essential in differentiating the nature of transport channels of TIs\cite{IGarate2012}. For each two dimensional conducting channel, a value of $\alpha=0.5$ is expected, and if there are two independent 2D conducting channels, the prefactors $\alpha$ add up. An effective dephasing length $l_{\varphi}$ will replace the phase relaxation length of the individual channel\cite{IGarate2012,JLee2012}. Fitting magnetoconductance data at different $V_{bg}$ to equation~\textcolor{blue}{(\ref{HLN})}, we observe a modulation of the prefactor $\alpha$ and of the dephasing length $\l_{\varphi}$ as presented in Fig.~\ref{fig:WALbehaviour}\textcolor{blue}{(b)}. The values of $\alpha$ drop from $\sim -0.5$ at high, positive $V_{bg}$ (region I), to a value of $\sim -1$ in the intermediate regime at $V_{bg}\simeq 50\,$V (region II), and saturate at a value slightly smaller than $ -1$ for negative back-gate voltages (region III). The same trend is observed in the $V_{bg}$ dependence of $\l_{\varphi}$. For large gate voltage, $V_{bg}=+200~V$, the phase coherence is long, $l_\varphi=1.35~\mu$m, but it reduces to $\l_{\varphi}\simeq 292$ nm when $V_{bg}=0\,$V and flattens out for negative gate voltages (region III). The extracted $\l_{\varphi}$ is larger than the
thickness of the sample (15 nm), indicating that the bulk channels, if present, are 2D with regard to WAL and will be contributing to the measured WAL signal when $E_f$ is located in the bulk band as pointed out previously for magnetotransport data of Bi$_2$Te$_3$ samples \cite{Chiu2013,HTHe2011}.

Schematic band diagrams are depicted in Fig.~\ref{fig:WALbehaviour}\textcolor{blue}{(c)} illustrating the evolution of the shift of the Fermi level for the three different back-gate voltage regimes. Our data suggest that at low temperature, we can use the back-gate to tune the Fermi level all the way from the BCB to the topological surface states; and then to the BVB. In most cases, WAL data in TI materials are governed by the competition between phase coherence time $\tau_{\varphi}$ and the surface-to-bulk scattering time $\tau_{SB}$ \cite{Kim2013,Steinberg2011,MLang2013,MBrahlek2014,JTian2014}. 
When the carrier density is high, and the Fermi level ($E_\text{f}$) is located inside the bulk conduction band (BCB, Fig.~\ref{fig:WALbehaviour}\textcolor{blue}{(c)}, right-hand side). If substantial surface-to-bulk scattering is present, the effective phase coherence time will be much larger than the surface-to-bulk scattering time, $\tau_{\varphi} \gg \tau_{SB}$, resulting in a single effective 2D transport channel with a prefactor of 0.5 to WAL. In the opposite limit, $\tau_{\varphi} \ll \tau_{SB}$, surface states and bulk are decoupled. Each region constitutes a separate coherent channel leading to a correction to the sample conduction of the order of $e^2/h$. Whereas the topological surface states contribute with a negative sign, a positive correction with $\alpha \approx +0.5$ is expected from the bulk conduction band when the Fermi level is close to the bottom of the band\cite{HZLu2014}. The individual contributions cancel partially, and we are left with a conductance correction $\approx -0.5 e^2/h$. 

After depleting the bulk carriers, region II (Fig.~\ref{fig:WALbehaviour}\textcolor{blue}{(c)}, center), the Fermi level $E_\text{f}$ crosses only the topological surface states. The two surfaces are decoupled, and we expect WAL with $\alpha = -1$, which is consistent with the observed value. Similar bottom-gating WAL results with $\alpha$ tuned from -0.5 to -1 has been reported in Bi$_2$Se$_3$ thin films grown on STO [111] \cite{Zhang2011,JChen2011} and Bi$_2$Te$_3$ nanoribbons placed on a STO-100 substrate\cite{AJauregui2015}. It should be pointed out that this scenario demands that both conduction channels have nearly identical dephasing fields, which requires good control of the Fermi level and surface morphology~\cite{JChen2011,JChen2010}. At negative $V_{bg}$, $\alpha$ saturates at a value close to unity $(\alpha \sim -1.1)$ indicating that the surfaces stay decoupled as we lower the Fermi level (region III in Fig.~\ref{fig:WALbehaviour}\textcolor{blue}{(c)} left-hand side). 

\subsection{Back-gate dependence of the in-situ Te-capped Bi$_2$Te$_3$ film}
\begin{figure}[!t]
{\includegraphics[width=0.45\textwidth]{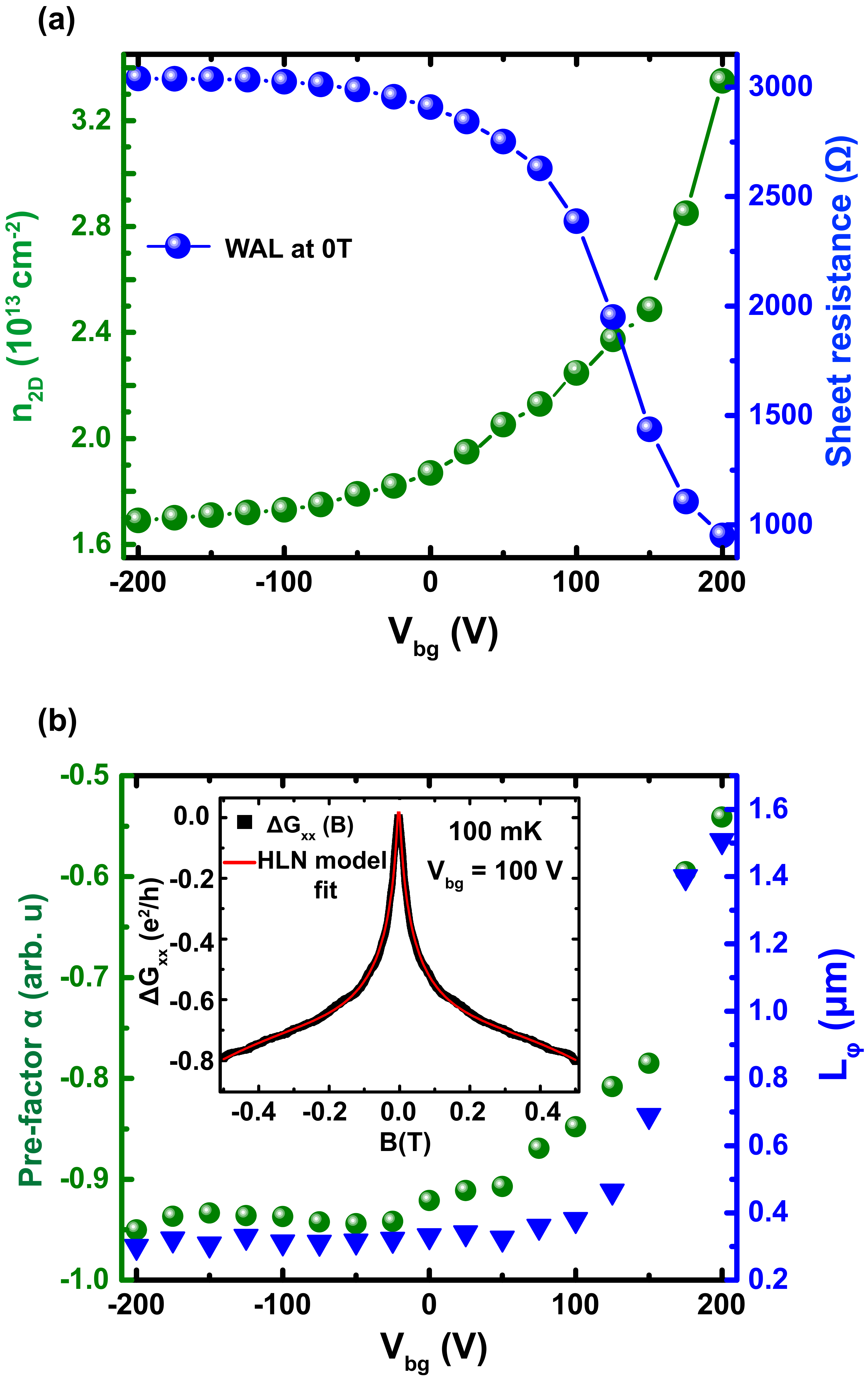}}  
  \caption{Transport characteristics of a Bi$_2$Te$_3$ film capped in-situ with a $6$ nm Te layer. (a) Back-gate voltage ($V_{bg}$) dependence of the 2D carrier density extracted from the zero-field slope of $R_\text{xy}(B)$ and of the sheet resistance. (b) Back-gate dependence of the WAL effect of an in-situ Te-capped Bi$_2$Te$_3$ film. The prefactor $\alpha$ and the dephasing length $l_{\varphi}$ as function of $V_{bg}$, extracted from best fits to equation \textcolor{blue}{(\ref{HLN})} as illustrated in the inset.}
  \label{fig:TransportTecappedFilms}
\end{figure}

In Figure~\ref{fig:TransportTecappedFilms}\textcolor{blue}{(a)}, we plot the carrier density  $n_\text{2D}$  and the sheet resistance of the Te-capped sample. The sheet carrier density values were extracted from the linear part of the low-field Hall resistance $R_{\text{xy}}(B)$. The prefactor $\alpha$ and the dephasing length $l_{\varphi}$ obtained from fit to equ.~\ref{HLN} are shown in Figure~\ref{fig:TransportTecappedFilms}\textcolor{blue}{(b)}
for different back-gate voltages. We observe a similar characteristic as in the Al$_2$O$_3$-capped Bi$_2$Te$_3$ film. The change in the total number of (mobile) charge carriers is $\Delta n_\text{2D} = 1.5\times 10^{13}$ cm$^{-2}$ (see Fig.~\ref{fig:TransportTecappedFilms}\textcolor{blue}{(a)}). This variation in charge carriers is comparable to the values reported recently for bottom-gated Se-capped Bi$_2$Se$_3$ films grown on an amorphous SiO$_2$/Si substrate\cite{YHLiu2015}. Sweeping the back-gate, both the sheet resistance and the carrier concentration saturate at around $-50\,$V, the lowest carrier density being $n_\text{2D}\simeq  1.7\times 10^{13}$ cm$^{-2}$ at $-200\,$V. Although we do not observe a maximum in the sheet resistance or the ambipolar effect in the $R_{\text{xy}}(B)$ data within the gate voltage range, the sharp increase in the sheet resistance together with the corresponding sharp drop in 2D carrier density suggest that we move the Fermi level through the bulk band gap region of the Bi$_2$Te$_3$ band structure diagram. At these energies the transport properties are dominated by the surface states as the bulk conductivity is significantly reduced \cite{YHLiu2015,SHong2012,YSKim2011,DKim2013}. 
\begin{table}[!t]
\centering
\begin{tabular}{|m{1.75cm}|m{2cm}|m{1.8cm}|m{2.cm}|}
\hline
 \textbf{Sample} & \vtop{\hbox{\strut \textbf{Capped/}}\hbox{\strut \textbf{Uncapped}}}&$\mathbf{n_{2D}\,\text{(\textbf{cm}}^{-2})}$& \vtop{\hbox{\strut $\mathbf{\mu \, (\text{\textbf{cm}}^2/\text{\textbf{V.s}})}$}\hbox{\strut \textbf{at} $\mathbf{B=0\,}$\textbf{T}}}\\ \hline
 \vtop{\hbox{\strut Sample-I}\hbox{\strut STO [111]}}& Uncapped & 4.3$\times$10$^{13}$& 964  \\ \hline
 \vtop{\hbox{\strut Sample-II}\hbox{\strut STO [111]}} &  Uncapped& 6.2$\times$10$^{13}$& 915 \\ \hline
 \vtop{\hbox{\strut Sample-III}\hbox{\strut Al$_2$O$_\text{3}$ [0001]}}& \vtop{\hbox{\strut Capped}\hbox{\strut with Al$_2$O$_3$}} & 9.5$\times$10$^{12}$&1206  \\ \hline
 \vtop{\hbox{\strut Sample-IV}\hbox{\strut STO [111]}} &   \vtop{\hbox{\strut Capped}\hbox{\strut with Al$_2$O$_3$}}   & 3.8$\times$10$^{12}$&1594 \\ \hline
 \vtop{\hbox{\strut Sample-V}\hbox{\strut STO [111]}}& Uncapped& 3.8$\times$10$^{13}$&826 \\ \hline
 \vtop{\hbox{\strut Sample-VI}\hbox{\strut STO [111]}} &Uncapped& 2.8$\times$10$^{13}$&1023  \\ \hline
 \vtop{\hbox{\strut Sample-VII}\hbox{\strut STO [111]}} &Uncapped & 1.2$\times$10$^{14}$& 41 \\ \hline
 \vtop{\hbox{\strut Sample-VIII}\hbox{\strut STO [111]}}&  \vtop{\hbox{\strut Capped}\hbox{\strut with Te}} & 1.8$\times$10$^{13}$&612 \\ \hline
\end{tabular}
\caption{Comparison of the charge carrier densities and the mobilities at zero gate voltage. All capped and uncapped thin films are of the same thickness (15~nm), and they were grown under the same conditions. Samples IV and VIII refer to the Al$_2$O$_3$- and Te-capped samples, respectively, which are discussed in detail in the main text.}
\label{table1_2}
\end{table} 

 Fits of the magnetoconductance curves to equation \textcolor{blue}{(\ref{HLN})} (see inset Fig.~\ref{fig:TransportTecappedFilms}\textcolor{blue}{(b)}), were carried out for all gate voltages. We observe a distinct modulation of the prefactor $\alpha$ and the dephasing length $l_{\varphi}$ with $V_{bg}$ shown in Fig.~\ref{fig:TransportTecappedFilms}\textcolor{blue}{(b)}. The results agree qualitatively with the Al$_2$O$_3$-capped Bi$_2$Te$_3$ sample in that both, $\alpha$ and $l_{\varphi}$, decrease as $V_{bg}$ is reduced. Again,  $\alpha$ approaches $-1/2$ for $V_{bg}>150$~V suggesting a single conduction channel coupled by a conducting bulk or, alternatively, a weak localization correction of $\sim 0.5\ e^2/h$ to the conductance from a decoupled bulk conduction band \cite{HZLu2014}. As the bulk carriers are depleted, $\alpha \approx -1$, suggesting that each surface acts as an independent coherent transport channel with comparable dephasing fields. 

To asses the role of the capping layers in protecting the topological surface states of Bi$_2$Te$_3$ thin films, we compare transport characteristics from capped films to uncapped ones. We choose films of the same thickness (15~nm), which were grown under the same conditions. Table~\ref{table1_2} summarizes the extracted low field carrier densities and the corresponding zero magnetic field mobilities for the two capped samples (sample IV and VIII), we presented in this work, and of a capped Bi$_2$Te$_3$ film grown on sapphire (sample-III) together with five other uncapped Bi$_2$Te$_3$ films grown on STO. Clearly, we observe an improvement in the transport characteristics (i.e.,  lower intrinsic carrier densities while maintaining similar or high mobilities) for the in-situ Al$_2$O$_3$- and Te-capped  films. 

\section{Conclusion}\label{FourFive}
We demonstrated the effectiveness of in-situ capping for preserving topological surface states in bulk insulating Bi$_2$Te$_3$ thin films. For this purpose, we tested capping with Al$_2$O$_3$ and Te layers and found that both methods lead to a significant decrease in the measured carrier density by protecting the TI surfaces from adsorbates and other contaminations. We back-gated the samples using the SrTiO$_3$ substrate as a dielectric, and we found that we were able to vary the carrier densities by $\sim 1.5 - 2 \times 10^{13} \text{cm}^{-2}$ carriers in the gate voltage range of $\pm 200~\text{V}$. In samples with low intrinsic doping, this modulation is sufficient to move from a bulk-dominated transport regime to a bulk insulating sample with decoupled surface states, which is evidenced by a doubling of the weak antilocalization correction to the sample conductance. The ability to access TI surface states is an important step toward device applications in the field of quantum information and spintronics. We showed that thin film deposition techniques of TI materials with low intrinsic doping paired with protective capping of the topological surface and effective depletion of bulk carriers by back-gating provide a technological platform for large-scale integration of such electronic devices. 
\newline
\paragraph*{}
This work is supported financially by the Dutch Foundation for Fundamental Research on Matter (FOM), the Netherlands Organization for Scientific Research (NWO) and by the European Research Council (ERC).

\end{document}